\documentclass[12pt]{iopart}
\usepackage{iopams}
\usepackage{epsfig}

\def\ead#1{\vspace*{5pt}\address{E-mail: \mailto{#1}}}
\def\mailto#1{{\tt #1}}   
  
\begin{document}

\title[Determination of the longitudinal structure function $F_{L}$ at HERA]
{Determination of the longitudinal structure function \boldmath{$F_{L}$} at HERA}

\author{Nelly Gogitidze  
}

\address{Lebedev Physical Institute RAS, 117924 Moscow, Russia \\ and DESY, Notkestrasse 85, 22607 Hamburg, Germany}
\ead{nellyg@mail.desy.de}

\begin{abstract}
Recent results from the HERA experiment H1 on the longitudinal stucture 
function $F_{L}$ of the proton are presented. They include proton structure 
function analyses with particular emphasis on those kinematic regions which
are sensitive to $F_{L}$. All results can be consistently described within 
the framework of perturbative QCD.
\end{abstract}




\section{Introduction}

Deep inelastic lepton nucleon scattering (DIS) experiments have played a
key role in the understanding of hadronic matter. They have revealed the
structure of hadrons being made out of constituents and the interaction 
between the latter. 

In fixed target DIS experiments scaling violations have been observed, i.e. 
the variation at fixed values of Bjorken-$x$ of the structure functions with 
$Q^2$, the squared four-momentum transfer between lepton and nucleon. These 
scaling violations are well described by perturbative Quantum Chromodynamics
(pQCD). The $Q^2$ 
evolution of the proton structure function $F_2(x,Q^2)$ is related to the 
gluon momentum distribution in the proton, $xg(x,Q^2)$, and the strong 
interaction coupling constant, $\alpha_s$. Both, $\alpha_s$ and $xg(x,Q^2)$, 
can be determined with precision $e p$ DIS cross section data.
 
HERA offers the possibility to study the structure of the proton over a wide 
kinematic range of $x$ and $Q^2$, several orders of magnitude  larger 
than in the earlier  fixed target experiments. Already first measurements 
of $F_2$ at HERA, for $x \sim 10^{-3}$ and $Q^2 \sim$ 20 GeV$^2$, revealed 
a steep rise of $F_2(x,Q^2)$ towards low $x$ for fixed $Q^2$ \cite{ref1,ref2}.
Later measurements of $F_2$ by H1\cite{H1F296,H1F200,H1F201} 
and ZEUS\cite{ZEUSF296,ZEUSF201} 
in a wide kinematic range, for $Q^2$ values from $1$ to 30000 GeV$^2$ and
for $3\cdot 10^{-5}<x<0.65$, have confirmed this trend.
The strong scaling violations observed at low $x$ are attributed to the high 
gluon density in the proton. The validity of the DGLAP evolution equation 
\cite{ref3}, which neglects higher-order terms \cite{ref4,ref5} proportional 
to $\alpha_s \cdot ln(1/x)$, is questionable in the low $x$ range and 
therefore has to be tested against the data. At sufficiently low $x$, 
non-linear gluon interaction effects have been considered in order to damp 
the rise of the cross section in accordance with unitarity requirements 
\cite{ref6}. 

The measurement of the longitudinal structure function $F_{L}(x,Q^2)$ is of 
great theoretical importance, since it may allow distinguishing between 
different models describing the QCD evolution at low $x$. In fact, the 
structure function measurements at HERA remain incomplete until the 
longitudinal structure function $F_L$ is actually measured.
The direct, classical $F_L$ measurement by substantial variation of the beam 
energies\cite{refBeam}
is yet to come at HERA. However, while waiting for this 
to happen in the HERA II period, the H1 collaboration have extracted 
experimental results on $F_L$ by measuring the cross section 
in a kinematic region where the $F_L$ 
contribution is substantial (the high $y$ region, see below)
 and comparing these measurements with the 
extrapolation of the $F_2$ measurements.
These results are described in this report.

\section{Neutral current cross section and the proton structure functions}

The cross section of the inclusive electron-proton DIS
neutral current process (Figure~1), $e p \rightarrow e X$, depends on the  
three independent variables $x$ and $Q^2$ and $s$, the center of mass energy 
squared. 
\begin{figure}[h]
\begin{center}
\epsfig{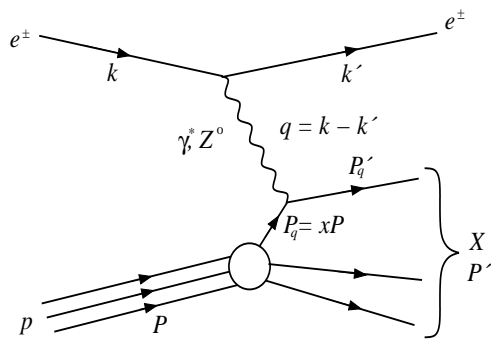}
\end{center}
\caption{\label{fig1} Feynman diagram of neutral current DIS.}
\end{figure}
$Q^2$ and $x$ are defined in terms of the four momenta of the incident and 
the scattered electron, $k$ and $k'$, and of the incident proton, $P$ as
\begin{equation}
Q^2 = -(k- k')^2 \qquad x = {Q^2\over 2P(k-k')} .
\end{equation}
Another important variable is the inelasticity
\begin{equation}
y = {P(k-k')\over P k}
\end{equation}
which is related to $x$, $Q^2$ and $s$ by $Q^2 = s x y$ \footnote[7]{ Here 
and in the following formulae the proton and electron masses are neglected.}. 

In the one-photon exchange approximation ($Z^{0}$ exchange neglected:
$Q^2 \ll m^2_{Z^{0}}$) the cross section of the 
$e p \rightarrow e X$ process can be expressed as
\begin{equation}
d \sigma \sim L_{\mu \nu} W^{\mu \nu}.
\end{equation}
Here $L_{\mu \nu}$ denotes the leptonic tensor, describing the interaction
between the electron and the virtual photon. In QED the lepton-photon vertex 
is well defined and therefore $L_{\mu \nu}$
can be calculated exactly. The hadronic tensor $W^{\mu \nu}$ corresponds to
the photon-proton interaction and is not calculable. However, using 
Lorenz invariance and current conservation the hadronic tensor can be 
reduced to only two functions related to the structure of the proton. 
The double differential DIS cross section can then be written as
\begin{equation}
{d^2 \sigma\over {dx dQ^2}} = {2 \pi \alpha^2\over {x Q^4}} (Y_{+} F_2(x,Q^2) - y^2 F_L(x,Q^2)) \qquad Y_{+} = 1 + (1-y)^2 .
\end{equation} 

One can consider the $e p$ scattering process as the interaction of a flux
of virtual photons \cite{refHand} with the proton. The DIS cross section can 
then be written as
\begin{equation}
{d^2 \sigma\over {dx dQ^2}} = {\Gamma (y) (\sigma_{T} + \epsilon (y) \sigma_{L})},
\end{equation} 
where $\Gamma (y) = Y_{+} \alpha / (2 \pi Q^2 x)$ stands for the flux factor,
$\epsilon (y) = 2 (1-y) / Y_{+}$ defines the photon polarization and 
$\sigma_{T}, \sigma_{L}$ correspond to the cross sections of the interaction
of transverse and longitudinally polarized photons, respectively. These cross
sections are related to the structure functions as 
\begin{equation}
\eqalign{
F_2(x,Q^2) = {Q^2\over {4 \pi^2 \alpha}} (\sigma_{T}(x,Q^2) + \sigma_{L}(x,Q^2)) \\
F_L(x,Q^2) = {Q^2\over {4 \pi^2 \alpha}} \sigma_{L}(x,Q^2) .}   
\end{equation} 
The longitudinal structure function $F_{L}$ is directly proportional to
$\sigma_{L}(x,Q^2)$. Due to the positivity of the cross sections, the two 
structure functions $F_2$ and $F_L$ obey the relation 
\begin{equation}
0 \le F_L \le F_2.
\end{equation}
 
The ``reduced'' cross section is defined as
\begin{equation}
\sigma_{r} \equiv F_2(x,Q^2) - {y^2\over {Y_{+}}} {F_L(x,Q^2)} . 
\end{equation}
Since the contribution of the longitudinal structure
function $F_L$ to the cross section can be sizeable only at large 
values of $y$, in a large kinematic range the relation 
$\sigma_{r} \approx F_2$ holds to a very good approximation. 

In the Quark Parton Model (QPM) \cite{refQPM}  the structure function $F_2$
can be expressed as a sum of the quark-antiquark momentum distributions
$x q_{i}(x)$ weighted with the square of the quark electric charges $e_{i}$
\begin{equation}
F_2(x) = \sum_{i} e^2_{i} x (q_{i}(x) + \bar q_{i}(x)) . 
\end{equation}
For spin 1/2 partons QPM also predicts $\sigma_{L} = 0$, which
leads to the so-called Callan-Gross relation \cite{refCG}: $F_{L}(x) = 0$.

The naive QPM has to be modified in QCD as quarks 
interact through gluons, and can radiate gluons. Radiated gluons in turn can
split into quark-antiquark pairs (``sea quarks'') or gluons. The gluon 
radiation
results in a transverse momentum component of the quarks. Consequently, quarks
can also couple to longitudinally polarized photons and the Callan-Gross
relation is no longer satisfied exactly. Thus, in QCD the longitudinal 
structure function
$F_L$ is non-zero. Due to its origin, $F_L$ is directly dependent on the 
gluon distribution in the proton and therefore the measurement of $F_L$
provides a sensitive test of perturbative QCD.

\section{Experimental procedure}
 
\subsection{Kinematic reconstruction}

At HERA 27.5 GeV positrons collided with 820 GeV protons. For the last 
two years the proton beam energy was increased to 920 GeV, 
and consequently the center of mass energy from 300 to 319 GeV.

The kinematic variables $y$ (or $x$) and $Q^2$ can be reconstructed  
either from the energy and the polar angle of the scattered positron, or from
the energy and the polar angle of the hadronic final state, or using a 
combination of these energies and polar angles. The overconstrained kinematics
allows cross checks, as well as the choice of the method with the best 
experimental precision for a given kinematic range.   

In the ``electron method'' the event kinematics is reconstructed using the
energy $E'_{e}$ and the polar angle $\theta_{e}$ of the scattered positron
\begin{equation}
Q^2_{e} = {E'^2_{e}sin^2 \theta_{e}\over {1 - y_{e}}} \qquad y_{e} = 1 - {E'_{e}\over {E_{e}}} sin^2 (\theta_{e}/2) . 
\end{equation}
While the electron method is accurate at large values of $y$, corresponding 
to low energies $E'_{e}$, the resolution rapidly degrades for 
$y_{e} \rightarrow 0$ where $E'_{e}$ approaches the positron beam energy 
$E_{e}$. The kinematic variables at low $y$ are reconstructed using also 
information from the hadronic final state. The inelasticity $y$ can also be 
determined as \cite{refYh}
\begin{equation}
y_{h} = {\Sigma_{i}(E_{i} - p_{z,i})\over {2 E_{e}}} = {\Sigma_{h}\over {2E_{e}}},
\end{equation}
where $E_{i}$ and $p_{z,i}$ are energy and longitudinal momentum components
of the particle~$i$ in the hadronic final state. In the ``$\Sigma$~method''
\cite{refSigma} $Q^2$ and $y$ are reconstructed as 
\begin{equation}
Q^2_{\Sigma} = {E'^2_{e} sin^2 \theta_{e}\over {1 - y_{\Sigma}}} \qquad y_{\Sigma} = {y_{h}\over {1 + y_{h} - y_{e}}} = {\Sigma_{h}\over {\Sigma_{h} + E'_{e}(1 - cos \theta_{e})}} .
\end{equation}
The $\Sigma$~method has good resolution also at low $y$, and is insensitive 
to the initial state radiation. The cross section is extracted using a
combination of the electron method and the $\Sigma$~method.

\subsection{H1 detector}

The H1 detector \cite{refDet} combines tracking in a solenoidal magnetic field
of 1.15 T with nearly hermetic calorimetry to investigate high energy $e p$ 
interactions at HERA. For polar angles $\theta_{e} > 153^{o}$ the scattered 
positron energy $E'_{e}$ is measured in the backward electromagnetic lead-fibre
scintillator calorimeter (SPACAL) \cite{refSpacal} which has an integrated 
timing function to veto proton beam induced background interactions. Here
$\theta$ is defined with respect to the proton beam direction. Identification 
of the scattered positron is improved and the polar angle measured with a
backward drift chamber (BDC), situated in front of the SPACAL, and with the new
backward silicon detector (BST) \cite{refBst}. The BST consists of four 
detector planes, arranged perpendicular to the beam axis and equipped 
with 16 wedge shaped, double metal silicon strip detectors. The BST measures
the polar angle of the crossing tracks with an internal resolution of about
0.2 mrad for $\theta$ between 172$^{o}$ and 177$^{o}$. For polar angle
$4^{o} < \theta < 154^{o}$ the scatterd positron is measured in the LAr 
calorimeter.

The hadronic final state is reconstructed using the Liquid Argon (LAr)
calorimeter, tracking detectors and the SPACAL. The interaction vertex is 
determined with the central drift chambers, the jet chamber CJC and two layers
of $z$ drift chambers, mostly using the hadronic final state particles. This
vertex determination is complemented by the inner proportional chamber CIP ,
for $167^{o} < \theta_{e} < 171^{o}$, and by BST, for 
$171^{o} < \theta_{e} < 176.5^{o}$.

The luminosity is determined with a precision of 1.5\% using the small-angle 
bremsstrahlung process $e p \rightarrow e p \gamma$ \cite{refLumi}. The final
state photon and positron, scattered at very low $Q^2$, can be detected in 
calorimeters (``photon and electron taggers'') which are situated close to
the beam pipe at distances of 103 m and 33 m from the interaction point in
the positron beam direction.

\subsection{Event selection and simulation}

The event selection criteria are slightly different in the low $Q^2$ and in 
the high $Q^2$ ranges. In the low $Q^2$ range the scattered positron is 
identified in the SPACAL. In the high $Q^2$ range the positron is seen in the 
LAr calorimeter as that cluster of maximum transverse energy, for which the 
requirements on the cluster shape are satisfied. The difference is then mainly
in the cluster definition and in the track validation. 
In the LAr cluster-track validation we use CJC tracks for $\theta >35^{o}$, 
in the SPACAL cluster case we use CJC, BDC and BST tracks, where appropriate.  

Longitudinal momentum conservation in neutral current DIS events gives the
constraint that $E - p_{z}$, summed over the final state particles, is about
equal to $2E_{e}$. In events with initial state radiation the radiative photon
may carry a significant fraction of the $E - p_{z}$ sum. Such events are thus 
suppressed by requiring $E - p_{z} >$ 35 GeV. 

Distributions of the selected events were compared with simulations of deep
inelastic scattering, photoproduction and Compton scattering. The simulated
Monte Carlo events were submitted to the same reconstruction and analysis chain
as the real data. The track detection efficiencies were determined from the
data utilizing the redundancy of the central and backward tracking detector 
system. Agreement at the few per cent level was reached between experiment
and simulation for the large set of technical and physics distributions 
studied. 
 
\section{The longitudinal structure function $F_{L}(x,Q^{2})$}

\subsection{Photoproduction background estimate}

For the measurement of the longitudinal structure function it is essential to 
reach the highest possible values of $y$. This requires the identification of
scattered positrons with energies down to only a few GeV 
as well as the efficient 
rejection of photoproduction background events in which low energy deposits in
the SPACAL can mimic the signature of a deep inelastically scattered positron.

The main part of the background is due to photons from $\pi^{o} \rightarrow 
\gamma \gamma$ decays. In the low $Q^2$ region a sizeable fraction of this
background can be removed by requiring a track signal in the BST. The 
remaining background is due to photon conversion and showering in the
passive material in front of the detectors, possible overlap of $\pi^{o}$ 
decays with charged tracks, 
and misidentified charged pions. It is subtracted bin by bin using the 
PHOJET \cite{refPhojet} MC simulation. The photoproduction background can be 
estimated experimentally using tagged events in which the scattered positron 
is detected in the electron tagger.

At $Q^2$ above 10 GeV$^2$, for $y < 0.75$ the photoproduction background is 
subtracted using the PHOJET simulation. For $y$ values above 0.75 
experimental information is used by employing the charge assignment of 
central tracks associated with SPACAL energy clusters. This allows the energy 
range to be extended down to $E'_{e} =$3 GeV, corresponding to $y \le 0.89$. 
The candidates with negative charge are taken to represent the background in 
the positron data sample. This statistical subtraction procedure requires the 
study of any process which may cause a charge asymmetry. Suchs asymmetry is 
e.g. due to the antiproton interaction cross section exceeding that for proton 
interactions at low energies \cite{refAsym}. Annihilation leads to larger 
energy deposits in SPACAL than proton interactions which introduces an 
asymmetry for low energies above a given threshold. This charge asymmetry can 
be measured using tagged photoproduction events which fulfill the DIS event 
selection criteria and is taken into account in the measurement of the 
positron DIS cross section at high $y$. 

As already stressed above, 
the detailed understanding of the data down to 3 GeV scattered positron 
energy is essential for the measurement of the DIS cross sections at high $y$,
enabling the determination of $F_{L}$. Control plots illustrating the
cross section measurements at high $y$ are shown in Figures 2 and 3.
\begin{figure}[t]
\begin{center}
\epsfig{file=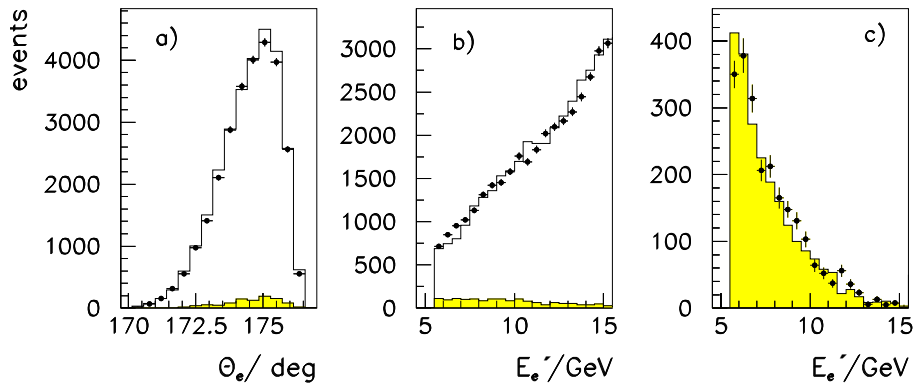,width=12.0cm,height=5.0cm}
\end{center}
\caption{\label{fig2} Distributions illustrating the cross-section
measurement at high $y$ ($0.46<y<0.82$) and 
$2~<~Q^2~<~5$~GeV$^2$ for events in the BST acceptance range.
DIS event distributions of a) the polar 
angle and b) the SPACAL energy of the scattered positron. 
c) SPACAL energy distribution for tagged photoproduction events 
fulfilling the DIS event selection criteria, apart from the $E - p_{z}$ 
requirement. Solid points: H1 data; shaded histograms: simulation of 
photoproduction events; open histograms: added distributions of simulated DIS 
and photoproduction events.}
\end{figure}
\begin{figure}[ht]
\begin{center}
\epsfig{file=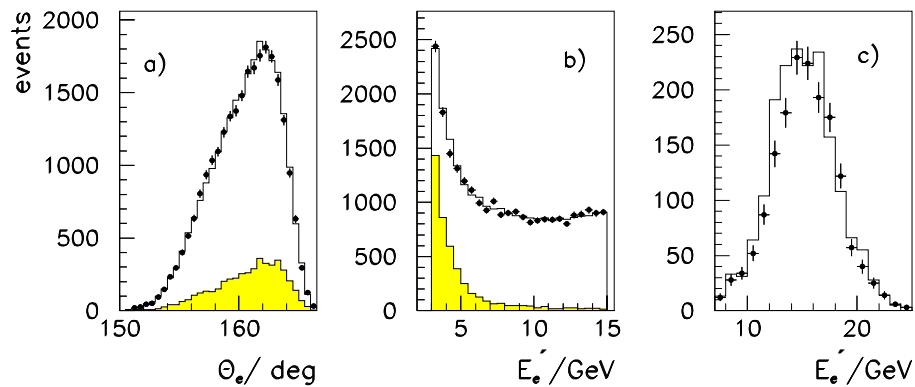,width=12.0cm,height=5.0cm}
\end{center}
\caption{\label{fig3} Distributions illustrating the cross-section measurement
at high $y$ ($0.46<y<0.89$) and $10~<~Q^2~<~35$~GeV$^2$. a) Polar 
angle and b) SPACAL energy distributions before subtraction of the
photoproduction background using the charge measurement by the CJC. Solid 
points: data with positive charge assignment. Shaded histogram: data with
negative charge assignment. Open histogram: sum of data with negative charge 
assignment and DIS event simulation, normalised to the integrated luminosity
of the data.
c) Spectrum of energy measured in the electron tagger for DIS
candidate events with a linked track of either positive charge (solid points)
or negative charge (histogram).}
\end{figure}

\subsection{Cross section at large $y$}

\begin{figure}[b]
\begin{center}
\epsfig{file=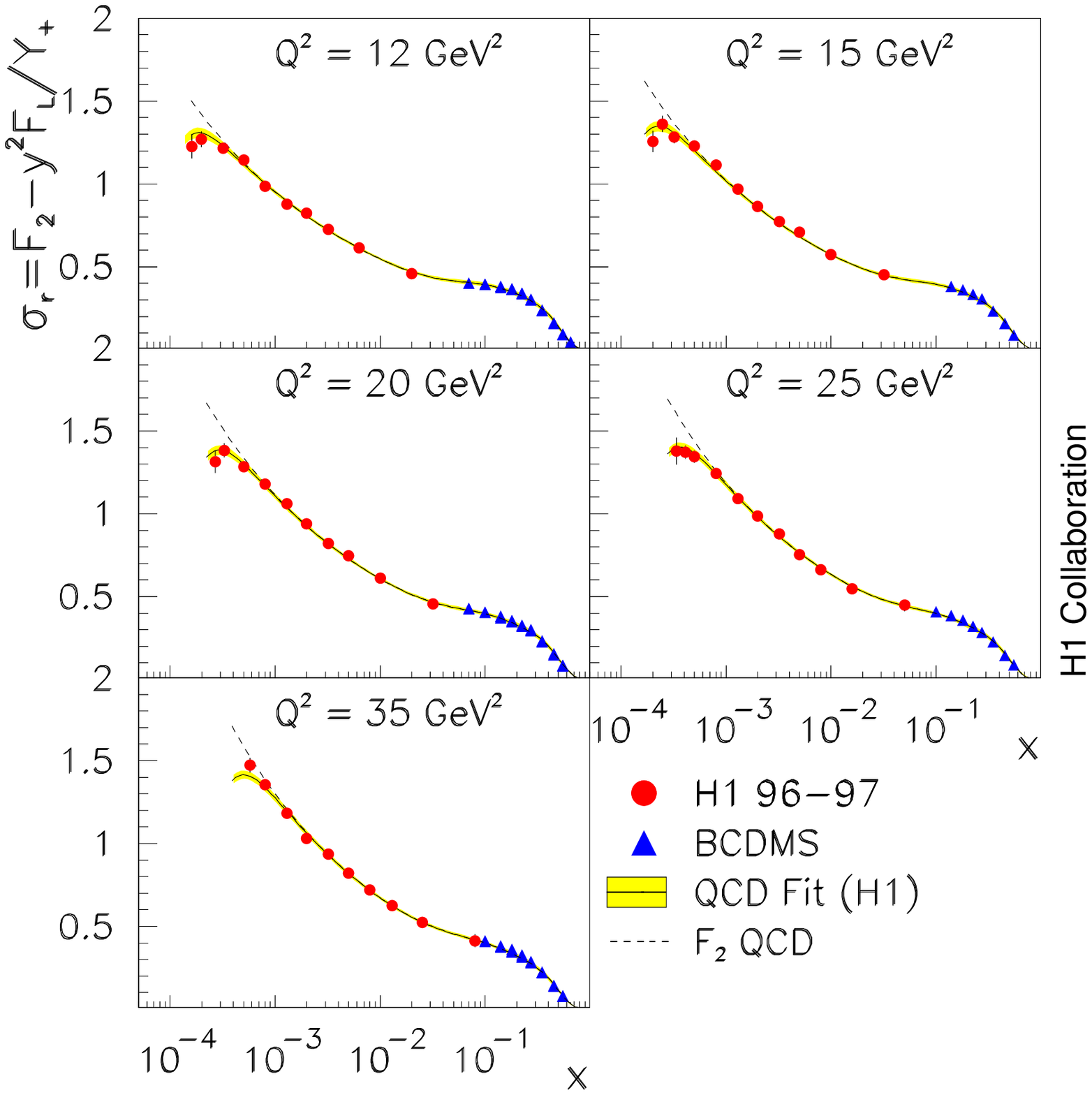,width=12.0cm,height=10.0cm}
\end{center}
\caption{\label{fig4}Measurement of the reduced DIS cross section
(closed points). Triangles represent data from the BCDMS muon-proton 
scattering experiment. The curves represent a NLO QCD fit to the H1 data 
alone, using data with $y < 0.35$ and $Q^2 \ge 3.5$ GeV$^2$. The dashed curves
show the $F_2$ structure function as determined with this fit. The error bands
represent the experimental and model uncertainty of the QCD fit.}
\end{figure}

The cross section is determined by converting the measured number of events,
after background subtraction, in bin averaged cross sections using
acceptance calculations from the Monte Carlo sample.  The precision for this 
measurement is dominated by systematic uncertainties of typically 3\%, 
extending to about 7\% at the edges of the covered $y$ range.

The measured reduced cross section $\sigma_{r}$ is shown in Figures 4 and 5 
for two $e^{+} p$ data samples, taken in 1996/97 and 1999 with 820 GeV and 
920 GeV proton beam energy, respectively. The cross section rises towards 
low $x$, turning over at $x$ values corresponding to $y \sim $0.6. This 
behaviour, for $Q^2 \ge $ 3.5 GeV$^2$, is well described by the NLO QCD fit 
\cite{refMethod1} to the 820 GeV data. Note that this fit used H1 data with 
a minimum $Q^2_{min}$ = 3.5 GeV$^2$. The extrapolation of the fit to $Q^2$ 
values lower than 2.5 GeV$^2$ falls below the measured data.
\begin{figure}
\vspace*{-1.0cm}
\begin{center}
\epsfig{file=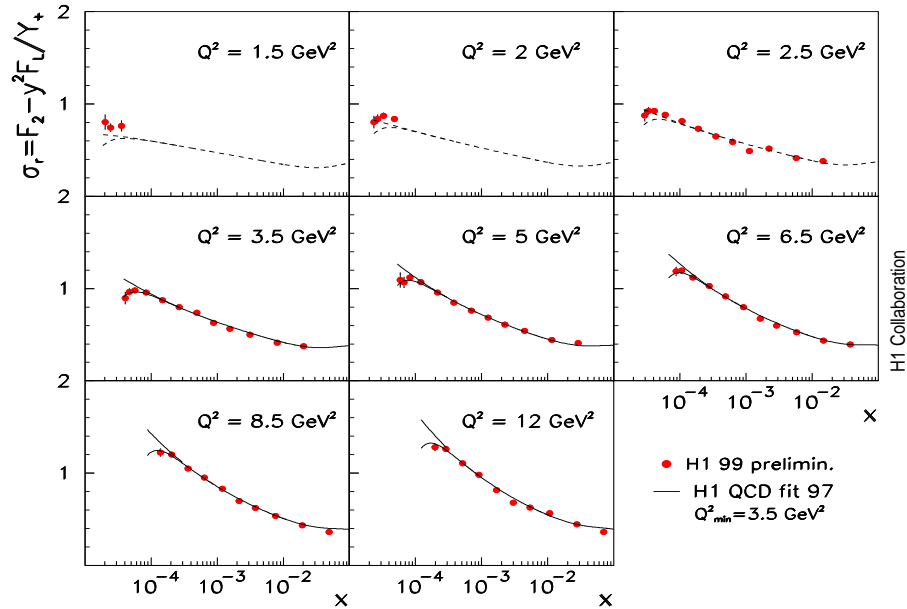,width=12.0cm,height=10.0cm}
\end{center}
\caption{\label{fig5} Measurement of the reduced DIS cross section (closed 
points) from the $e^{+} p$ data taken in 1999 at 920 GeV proton energy 
\cite{data99}. The 
curves represent a NLO QCD fit to the H1 1996/97 data, which was limited to 
$Q^2 \ge 3.5$ GeV$^2$. The dashed curves show the extrapolation of this fit
below $Q^2_{min}$. For each $Q^2$ bin the lower curves (turning at low x)
represent the reduced cross section $\sigma_{r}$ and 
the upper curves represent
the structure function $F_2$, which rises towards low $x$. For large $x$,
corresponding to low $y$, $\sigma_{r} \simeq F_2$.}
\end{figure}

\subsection{Determination of $F_{L}$}

The extraction of the longitudinal structure function $F_{L}$ is based on the 
reduced cross section (8), which depends on $F_{2}(x,Q^2)$
and $F_{L}(x,Q^2)$. The effect of $F_{L}$ is enhanced with $y^2$, and the 
reduced cross section $\sigma_{r}$ tends to $F_2 - F_{L}$ as 
$y \rightarrow 1$. For the extraction of the longitudinal structure function 
from the inclusive cross section the data at large $y$ are very important.
An important advantage of HERA, compared to fixed target DIS lepton-nucleon
experiments, is the wide range of $y$ values covered. This allows the 
behaviour of $F_2$ at low $y$ to be determined reliably and to be extrapolated
into the region of high $y$. 

Two methods are used by H1 to perform the extraction of the longitudinal 
structure function:

For $Q^2 >$ 10 GeV$^2$ the ``extrapolation method'', introduced in 
\cite{refMethod1}, is used. A NLO DGLAP QCD fit is used to
extrapolate $F_2$ into the high $y$ region. This fit uses only H1 data in the 
restricted kinematic range $y < 0.35$ and $Q^2 \ge 3.5$ GeV$^2$. In Figure 4
the fit is compared with the measured cross section for $Q^2$ bins above  
10 GeV$^2$, accessing the high $y$ region. The difference between the
measured $\sigma_{r}$ and the extrapolated $F_2$ is used to determined 
$F_{L}(x,Q^2)$. Systematic errors, which are common to the lower $y$ and
the larger $y$ region, are considered in the fit as described in \cite{refPZ}.
  
At low $Q^2 <$ 10 GeV$^2$, the behaviour of $F_2$ as a function of $\ln y$ is
used in a new extraction method \cite{H1F201}. This so called 
`` derivative method'' is based on the cross section derivative 
$(\partial \sigma_{r} / \partial \ln y)_{Q^2}$.

The derivative of the reduced cross section, taken at fixed $Q^2$, is given by
\begin{equation}
\left({\partial \sigma_{r}\over {\partial \ln y}}\right)_{Q^2} = \left({\partial F_2\over {\partial \ln y}}\right)_{Q^2} - F_{L} \cdot 2y^2 \cdot {{2 - y}\over {Y^2_{+}}} - {\partial F_{L}\over {\partial \ln y}} \cdot {y^2\over {Y_{+}}} .
\end{equation}
\begin{figure}[h]
\begin{center}
\epsfig{file=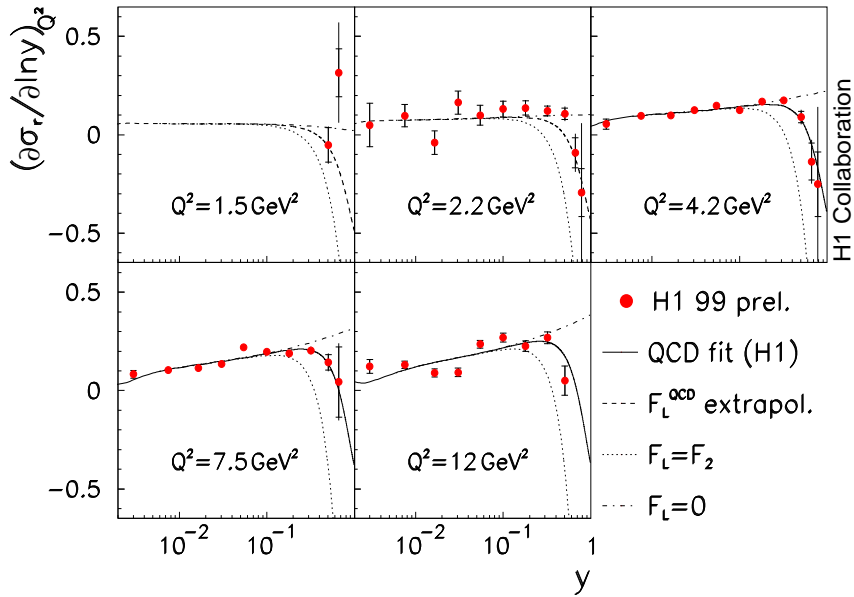,width=11.0cm,height=8.0cm}
\end{center}
\caption{\label{fig6}Measurement of the derivative $(\partial \sigma_{r} / 
\partial \ln y)_{Q^2}$. The curves represent the QCD fit result to the H1 
1996/97 data for $y <$ 0.35 and $Q^2 \ge$ 3.5 GeV$^2$ calculated with different
assumptions on $F_{L}$. The inner error bars represent the statistical errors 
and the total error bars the statistical and systematic errors, added in
quadrature.}
\end{figure}
For $y \rightarrow 1$ the cross section derivative tends to the limit 
$(\partial F_2 / \partial \ln y)_{Q^2} - 2 \cdot F_{L}$, neglecting the
contribution from the derivative of $F_{L}$.
At largest $y$ the $F_{L}$ contribution dominates the derivative of the 
reduced cross section $\sigma_{r}$. This is in contrast to the influence of
$F_{L}$ on $\sigma_{r}$ which is dominated by the contribution of $F_2$ for 
all $y$. A further advantage of the derivative method is that it can be applied
down to very low $Q^2 \simeq$ 1 GeV$^2$ where a QCD description of $F_2(x,Q^2)$
is complicated due to higher order and possible non-perturbative corrections.
The cross section derivatives are shown in Figure 6 as function of $y$. This 
measurement of $\partial \sigma_{r} / \partial \ln y$ was used to determine 
the longitudinal structure function. The assumption has been made that the 
derivative $\partial F_2 / \partial \ln y$ is a linear function of $\ln y$ 
up to large $y$. In each $Q^2$ bin straight line fits were made to the 
derivative data for $y <$ 0.3. The line fits describe the data very well and
the extrapolation of the straight line was taken to represent the contribution 
of $F_2$ at high $y$. The extrapolations were compared with the values 
obtained from QCD and good agreement was found. The uncertainties of the 
straight line extrapolation were included into the systematic errors of the 
measurement. The small contribution of $\partial F_{L} / \partial \ln y$ to 
the derivative was corrected for by using NLO QCD and the size of the full 
correction was added to the overall error of the measured $F_{L}$.   

\subsection{Results}

\begin{figure}[h]
\begin{center}
\epsfig{file=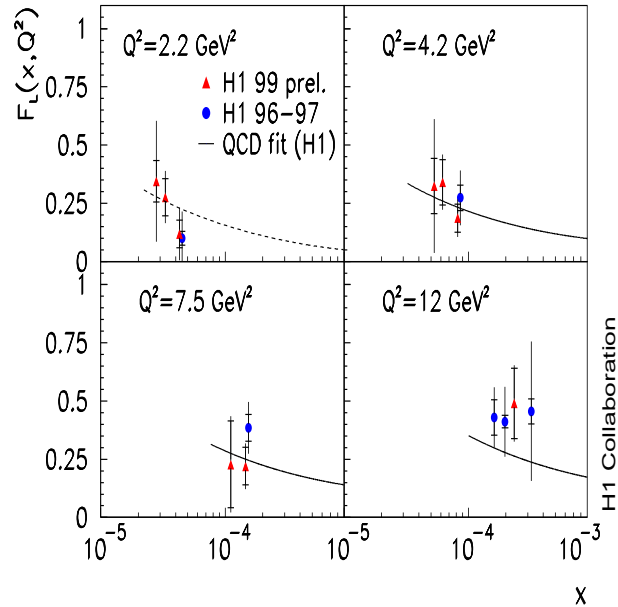,width=8.0cm,height=8.0cm}
\end{center}
\caption{\label{fig7}The longitudinal structure function $F_{L}(x,Q^2)$ as a
function of $x$ in bins of $Q^2$, determined with the derivative method from 
the $e^{+} p$ data taken in 1996/97 (points) and in 1999 (triangles). The 
inner error bars represent the statistical error and the total error bars the 
statistical and systematic errors, added in quadrature. The curves represent 
the QCD fit result to the H1 1996/97 data for $y <$ 0.35.}
\end{figure} 
The longitudinal structure function $F_{L}(x,Q^2)$ determined by the derivative
method is shown in Figure 7. The data extend into the lower $x$ range and are 
consistent with previous H1 results. The curves represent NLO QCD calculations 
of $F_{L}$ based on a fit to the 1996/97 data for $y <$ 0.35, i.e. in a region 
where $F_2$ can be measured practically independently of $F_{L}$.

In Figure 8 the determination of $F_{L}(x,Q^2)$ at high $Q^2$ is shown 
for $e^{+} p$ and $e^{-} p$ data sets. The extrapolation method is used.
Both data sets are consistent and
in agreement with the QCD fit. The possible extreme values for $F_{L}$
($F_{L} = 0$ and $F_{L} = F_2$) are shown for comparison. 
\begin{figure}[h]
\begin{center}
\epsfig{file=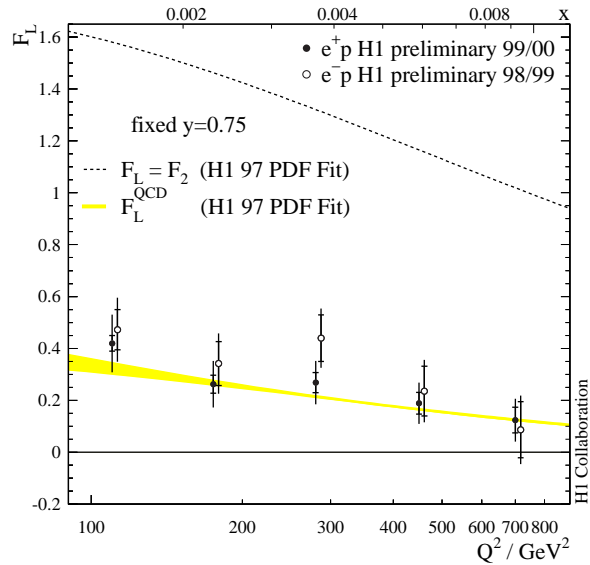,width=8.0cm,height=8.0cm}
\end{center}
\caption{\label{fig8}The longitudinal structure function $F_{L}$ as a function 
of $Q^2$ at fixed $y = $0.75 for the $e^{+} p$ and $e^{-} p$ data. 
The inner error bars represent the statistical error and the total error bars 
the statistical and systematic errors, added in quadrature. The curves 
represent the QCD fit result calculated with different assumptions on $F_{L}$.
The shaded band shows the expectation for $F_{L}$, and its uncertainty, from 
the QCD fit.}
\end{figure}

An overview of all current H1 data on $F_{L}(x,Q^2)$, from 
$Q^2$ = 2.2 GeV$^2$ to 700 GeV$^2$ is given in Figure 9. The data extend the
knowledge of the longitudinal  structure function into the region of low $x$,
much beyond the region of fixed target lepton-proton scattering 
experiments. The increase of $F_{L}(x,Q^2)$ towards low $x$ is consistent with
the NLO QCD calculation, reflecting the rise of the gluon momentum distribution
in this region. The values of $F_{L}(x,Q^2)$ are thus severely constrained
by the present data, unless there are deviations from the assumed 
extrapolation of $F_2$ into the region of large $y$ corresponding to the 
smallest $x$. A measurement of the $x$ dependence of $F_{L}(x,Q^2)$, 
independent of assumptions about the behaviour of $F_2$, can be performed 
with a variation of the proton beam energy at HERA \cite{refBeam}.
\begin{figure}[h]
\begin{center}
\epsfig{file=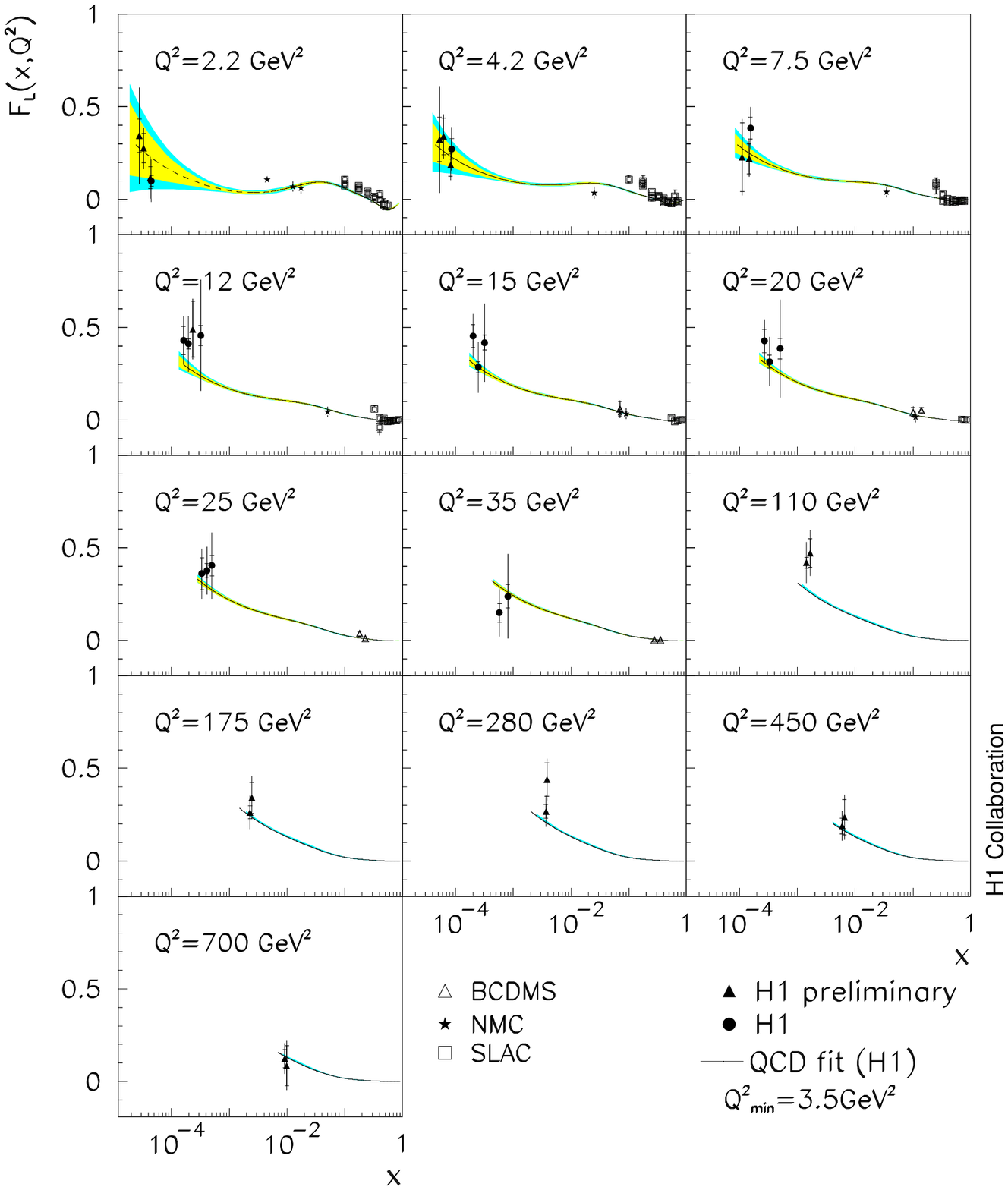,width=13.0cm,height=14.0cm}
\end{center}
\caption{\label{fig9}The longitudinal structure function $F_{L}(x,Q^2)$ as
a function of $x$ in bins of $Q^2$, obtained by H1 and by charged 
lepton-nucleon fixed target experiments. The inner error bars represent the 
statistical error and the total error bars the statistical and systematic 
errors, added in quadrature. The error bands are due to the experimental 
(inner) and model (outer) uncertainty of $F_{L}$ using the NLO QCD fit to 
the H1 1996/97 data for $y <$ 0.35 and $Q^2 \ge$ 3.5 GeV$^2$.}
\end{figure}

A NLO QCD fit has also been performed by the ZEUS collaboration 
\cite{refZEUS}. Within this fit $F_{L}$ was calculated, and found to be 
consistent with the H1 $F_{L}$ data.
 
Several theoretical models were compared to the H1 data. Predictions based
on the saturation model \cite{refGBW} are shown in Figure 10 together with the
H1 $F_{L}(x,Q^2)$ data points. At large $x$ and $Q^2$ values data are 
well described by the model, while at low $x$ and low $Q^2$ higher twist 
contributions are clearly needed to describe the $F_{L}(x,Q^2)$ distribution. 
\begin{figure}[h]
\begin{center}
\epsfig{file=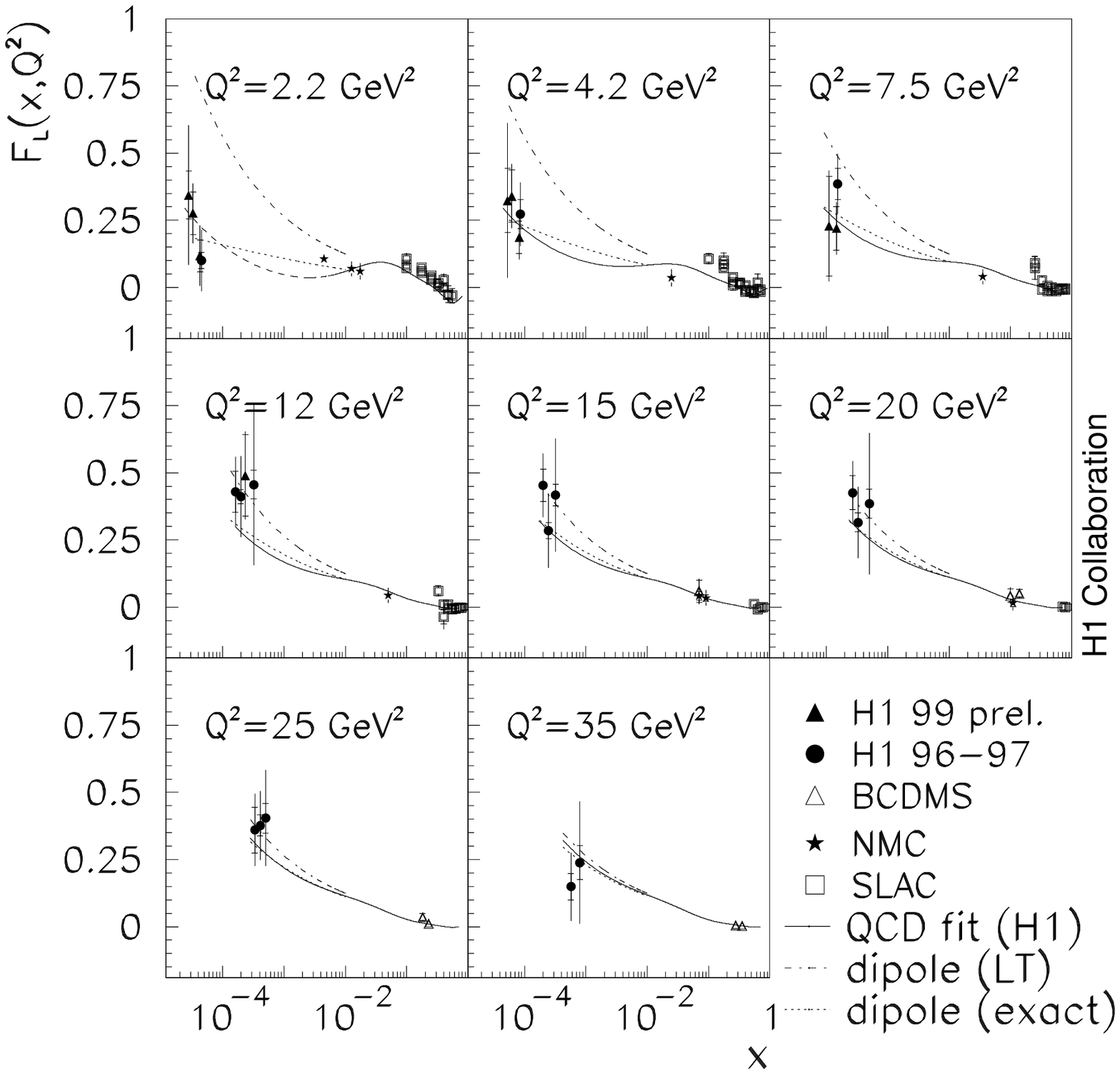,width=11.0cm,height=11.0cm}
\end{center}
\caption{\label{fig10}The longitudinal structure function $F_{L}(x,Q^2)$ as
a function of $x$ in bins of $Q^2$, obtained by H1 and by charged 
lepton-nucleon fixed target experiments. The inner error bars represent the 
statistical error and the total error bars the statistical and systematic 
errors, added in quadrature. The dash-dotted and dotted lines are fits using 
the saturation model of Golec-Biernat and W\"usthoff \cite{refGBW}, with 
higher twist contributions (exact) and without higher twists (LT).}
\end{figure}

In Figure 11 the H1 data are compared to
the Donnachie-Dosch model \cite{refDosch} 
predictions. This model is based on a dipole picture in which large 
dipoles couple to a soft pomeron and small dipoles couple to a hard pomeron. 
The parameters are fixed by proton-proton scattering data and by the measured 
proton structure function $F_2(x,Q^2)$. This simple approach is able to 
describe the $F_{L}$ data well. One can see that with increasing $Q^2$ the 
importance of the hard pomeron is increasing, and it is dominating at the high 
$Q^2$ values. 
\begin{figure}[h]
\begin{center}
\epsfig{file=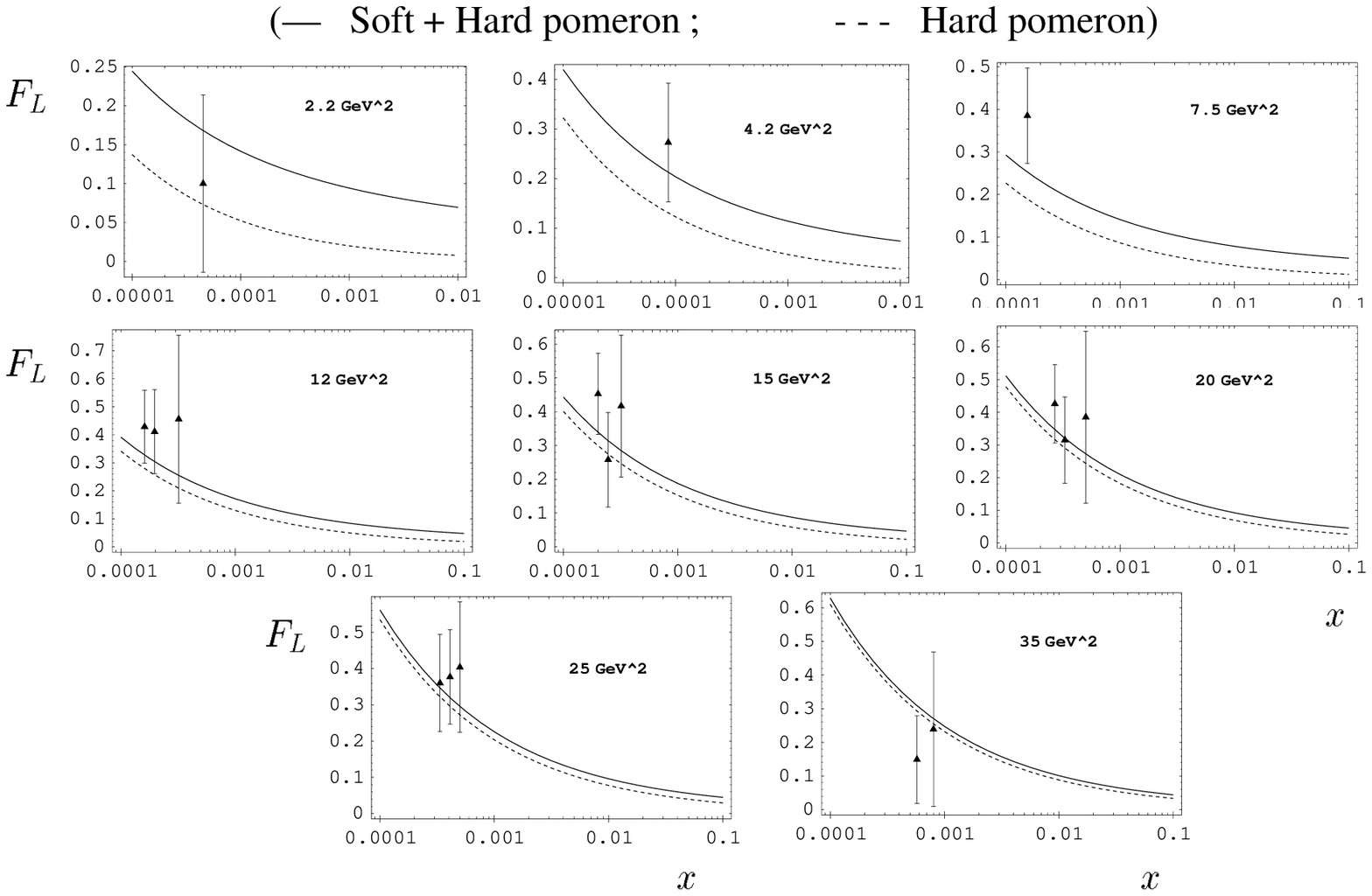,width=13.0cm,height=10.0cm}
\end{center}
\caption{\label{fig11}The longitudinal structure function $F_{L}$ as function
of $x$ in bins of $Q^2$, as measured by H1. Curves represent the 
Donnachie-Dosch model \cite{refDosch} predictions. The solid line is  
the full result and the dashed line is the hard pomeron contribution.}
\end{figure}
 
\section{Summary}

New results on the extraction of the 
longitudinal structure function $F_{L}(x,Q^2)$ are 
presented for a wide $Q^2$ range from 2.2 GeV$^2$ to 700 GeV$^2$ and for much
lower $x$ values ($3 \cdot 10^{-5} \le x$), compared to the fixed target 
experiments. The extraction results are possible 
due to the improved detectors in 
the H1 backward region, used for the identification and measurement of the 
scattered positrons at low $Q^2$ and high $y$.  

Two different methods are used to determine $F_{L}$. The partial derivative 
of the reduced cross section, $(\partial \sigma_{r} / \partial \ln y)_{Q^2}$, 
is used to extract $F_{L}$ at low $Q^2 <$ 10 GeV$^2$. At $Q^2 >$ 10 GeV$^2$
the difference between the measured reduced cross section $\sigma_{r}$ and
$F_2$, calculated from an extrapolation of a NLO QCD fit to low $y$ data, is
used for the $F_{L}$ determination. Thus the longitudinal structure function 
$F_{L}$ at low $x$ is determined more precisely than hitherto and in a larger
$Q^2$ range.

The observed rise of $F_{L}(x,Q^2)$ towards low $x$ is consistent with the 
NLO QCD calculations and reflects the rise of the gluon momentum distributions
in this kinematical region.

The data are also well described by calculations based on a 
colour dipole model and on a combination of the 
colour dipole with a two pomeron model.
 
\section*{Acknowledgments}

I would like to thank the organizers for the excellent organization and for
creating an inspiring conference atmosphere. The results presented here are
due to the common efforts and the success of the HERA machine group and of the
H1 Collaboration and it is a pleasure for me to thank all their members. 
Special thanks to J.~Dainton, D.~Eckstein, J.~Gayler, M.~Klein, K.~Long, 
J.~Olsson, and P.~Schleper for useful discussions, comments and help in 
preparing the talk.

\end{document}